\documentclass[12pt,preprint,flushrt]{aastex}
\usepackage{txfonts}
\usepackage{graphicx}

\begin{document}

\title{Studying the Imaging Characteristics of Ultra Violet Imaging Telescope (UVIT) through Numerical Simulations}

\author{Mudit K. Srivastava$^1$, Swapnil M. Prabhudesai$^2$, and Shyam N. Tandon$^3$}
\affil{Inter-University Centre for Astronomy and Astrophysics (IUCAA), Pune, India}
\email{
${}^1$mudit@iucaa.ernet.in\\
${}^2$smp@iucaa.ernet.in\\
${}^3$sntandon@iucaa.ernet.in
}

\begin{abstract} 
Ultra-Violet Imaging Telescope (UVIT) is one of the five payloads aboard the Indian Space Research Organization (ISRO)'s ASTROSAT space mission. The science objectives of UVIT are broad, extending from individual hot stars, star-forming regions to active galactic nuclei. Imaging performance of UVIT would depend on several factors in addition to the optics, e.g. resolution of the detectors, Satellite Drift and Jitter, image frame acquisition rate, sky background, source intensity etc. The use of intensified CMOS-imager based photon counting detectors in UVIT put their own complexity over reconstruction of the images. All these factors could lead to several systematic effects in the reconstructed images. 
\par
A study has been done through numerical simulations with artificial point sources and archival image of a galaxy from GALEX data archive, to explore the effects of all the above mentioned parameters on the reconstructed images. In particular the issues of angular resolution, photometric accuracy and photometric-nonlinearity associated with the intensified CMOS-imager based photon counting detectors have been investigated. The photon events in image frames are detected by three different centroid algorithms with some energy thresholds. Our results show that in presence of bright sources, reconstructed images from UVIT would suffer from photometric distortion in a complex way and the presence of overlapping photon events could lead to complex patterns near the bright sources. Further the angular resolution, photometric accuracy and distortion would depend on the values of various thresholds chosen to detect photon events. 
\end{abstract}


\maketitle


\section{Introduction}


ASTROSAT is a multi-wavelength space observatory to be launched by Indian Space Research Organization (ISRO) in 2009-2010. It consists of five astronomical payloads that would allow simultaneous multi-wavelengths observations from X-ray to Ultra-Violet (UV) of astronomical objects. The Ultra-Violet imaging telescope (UVIT) is one of these five payloads and is being developed with the aim of providing the flux calibrated images at a spatial resolution of $\sim$ 1.5 arc-second (arc-sec). UVIT records images simultaneously in three channels: Far Ultraviolet (FUV, 1300-1800 \AA), Near Ultraviolet (NUV, 1800-3000 \AA), and Visible (VIS, 3200-5300 \AA), simultaneously with $\sim$ 0.5 degree field of view. UVIT is configured as twin telescopes based on Ritchey-Chr\'{e}tien design with an aperture of $\sim$375 mm and focal length of $\sim$4750 mm: While one of these would make images in FUV while the other would be used for NUV and visible regime. Apart from the Optics and Opto-mechanical environment of UVIT, Detectors are the other critical elements that would govern the performance of UVIT imaging. Like other UV missions (FUSE, GALEX etc.), the UVIT detectors are also of photon counting nature based on Micro Channel Plate (MCP) image intensifiers Technology \citep{siegmund1999, jelinsky2003, sahnow2003}. However the read out scheme used in UVIT Detectors are completely different from the schemes used in FUSE and GALEX detectors. While the detectors used in FUSE and GALEX missions were read using a Double Delay Line and Cross Delay Line anode systems \citep{jelinsky2003, sahnow2003}, the UVIT detectors incorporates the readout system based on C-MOS devices. The similar readout systems based on CCD detectors have been discussed in the literature \citep{bellis1991, siegmund1999}. Due to this difference the imaging characteristics of the UVIT are expected to be very different from the other UV missions involving Photon Counting Detectors. 
\par
The UVIT Photon Counting Detectors are being developed through a collaboration of Canadian Space Agency (CSA), and ISRO. In this scheme a UV photon is detected through its footprint in form of a light splash on a CMOS detector. The exact coordinates of the photon would later be estimated through some centroid algorithms using the pixel values of the detector, to much higher resolution than one CMOS pixel. The details of photon-counting detectors to be used in UVIT are given in \cite{hutchings2007}. The CMOS detector consist of 512 X 512, 25 $\mu$m a side pixels. Each pixel extends $\sim$ 3 arc-sec $\times$ 3 arc-sec on the sky. The experimental studies done by \cite{hutchings2007} shows that one photon event produce a light splash which follows roughly a Gaussian distribution with Full Width at Half Maximum (FWHM) of $\sim$ 1.5 pixels. They have also reported the performance of several centroid algorithms that have been developed and applied to the laboratory image frames. The resolution of the detector is set by the gap between the photo-cathode and first MCP \citep{michel1997}. The UVIT detectors are designed with the gap of 0.1 - 0.15 mm to obtain a resolution of $\sim$ 1$''$. These intensified photon counting detector can either be run in a photon-counting mode with a very high gain (say $>$ 10,000 electrons/photon on CMOS ), which is the normal mode for the ultraviolet channels or in integrating mode with a low gain ( say 10-100 electrons/photon on CMOS), which is the normal mode for the Visible channel.
\par
Image reconstruction from the photon events centroid data is another area of concern. As estimated by ISRO the ASTROSAT Satellite would drift by roughly 0.2 arc-sec/sec during the observations. Therefore all the individual image frames would required to be corrected for this drift effect before co-adding them to create the final image. Hence the accuracy to which the satellite drift is determined and corrected, is another major factor that could affect the spatial resolution and photometric properties of the UVIT images. Also, apart from the performance of optics these properties would highly depend on the detector hardware design, choice of centroid algorithms, the effects of background and bias, frame acquisition rate etc. Further considering the nature of Photon Counting Detectors and readout process it is expected that the photometric non-linearity would be present in the reconstructed images.
\par
In this paper we report on the numerical simulations and the results that were performed to estimate the angular resolution and photometric properties of the UVIT system. The aim of this study is to explore the effects of various parameters and thresholds on the images generated by UVIT. Section-2 describes the process and results of the simulations to estimate the drift of the satellite. The process of UVIT data simulations, image reconstruction, different parameters/thresholds and related errors are discussed in Section-3. The various results regarding the photometric properties of the UVIT images for simulated point sources are presented in Section-4. In Section-5 UVIT images of some of the extended sky sources (adopted from archival GALEX and ACS data) are presented. The angular resolution of reconstructed UVIT images is discussed in Section-6. Finally Section-7 summarizes all the results.


\section{Simulation to Estimate Drift of ASTROSAT using Visible Channel of UVIT}


Though the optics and detectors of UVIT are designed to give a spatial resolution of $\sim$ 1.5 arc-sec, the drift of the satellite, while observing, is a major factor to overcome. In order to get sharp images, it is required to know the full time series of the satellite drift. To check if the images from the visible channel ( 3200 \AA - 5300 \AA) can be used to get this time series with the required accuracy, a simulation has been done to estimate accuracy of recovering drift aspect of satellite from the images in visible channel. The images from the visible channel, taken every $\sim$ 1 sec, can be used to track the drift aspect of the telescope, which slowly drifts with drift of the satellite. To estimate the performance of visible channel for drift estimation, stars fields, taken from ESO Hubble Space Telescope (HST) Guide Star Catalog are simulated through the visible channel. Typically a field of $\sim$ 29 arc-minute diameter is selected to match the field of UVIT. Simulations have been done for seven fields at different galactic latitudes. For each field, either all the stars brighter than magnitude 15 are taken or the brightest 100 stars are taken. The visible channel has an effective area of $\sim$ 25 cm$^2$, i.e. a star of magnitude 15 corresponds to $\sim$ 25 photons/sec.  
\par
\begin{figure}
\centering
\plotone{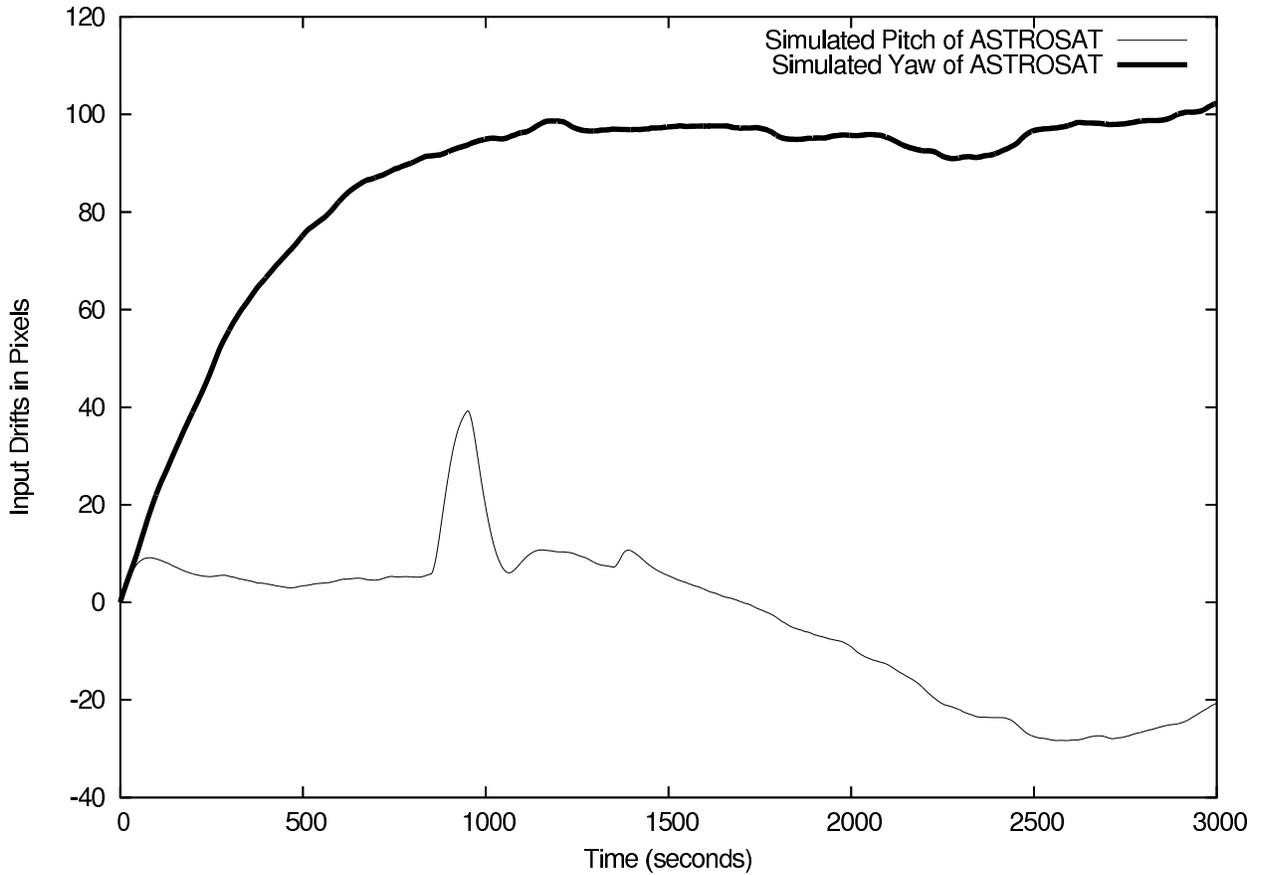}
\caption{Simulated Drift of the ASTROSAT Satellite as provided by ISRO Satellite Centre. This Drift is taken as input to simulate the process of drift estimation using visible channel of UVIT. The Drift in Pitch and Yaw directions are shown.}
\label{fig-drift}
\end{figure}

For each of these fields, the image is simulated for time interval of 1 second with a frame rate of 10 frames per second. The number of photons generated by each star in one second is estimated by Gaussian statistics with a variance equal to the average number of photons. In addition to the photons arising from stars, photons are introduced for the background. Later all these generated photons are randomly distributed in 10 image frames. To incorporate the Point Spread Function (PSF) of the optics, position of each photon is selected as per two dimensional Gaussian distribution, with an RMS (Root Mean Square) width of 0.3 pixel on either axis. Each photon is then assumed to produce, on average, 16 electrons following a Gaussian distribution with variance same as mean (In practice, the average number of electrons, for each photon, on the CMOS detector would be $\gg$ 16. The number 16 is used here to minimize the computation). The location of each electron in detector, is found by assuming a Gaussian distribution with an RMS width of 0.7 pixel on either axis. To allow for large gain, a multiplication factor of 10 is introduced to convert each of the electrons into Counts. All the 10 exposures are now added to get the frame for one second. The final simulated image frame is obtained by adding to each pixel a Gaussian read noise with an RMS value 10 Counts. Figure~\ref{fig-drift} shows the simulated cumulative satellite drift along Pitch and Yaw direction for 3000 seconds, by using the results of simulated drift rates (as obtained from a simulation done at ISRO Satellite Centre, ISRO). The effect of drift in the Roll is negligible on the CMOS detector as compare to drift in these two axes and hence not shown. Satellite drift is added to each of the image frames and so the image frames are drifted accordingly. Final image of the star field for 1 second is constructed by adding 10 consecutive frames. Various images of the star field is constructed at intervals of one second.
\par
To estimate the drift of the ASTROSAT, first 10 images (each of 1 second exposure) are added to obtain a reference-image. The centroids of stars in the reference-image are taken as standard, and the centroids in other one second exposure images are compared with this to find drift parameters, i.e. drifts in X and Y direction and the rotation, of the other images (corresponding to Pitch, Yaw and Roll motion of the satellite drift respectively). Further in order to minimize the errors due to noise, a local second order polynomial fit is made to the estimated drift parameters, for each interval of 20s. The fit is used to find a new value of the drift parameters at central point of the interval. The drifts parameters obtained thus are compared to the input values to get an estimate of the errors on these.
\par
\begin{figure}
\centering
\plotone{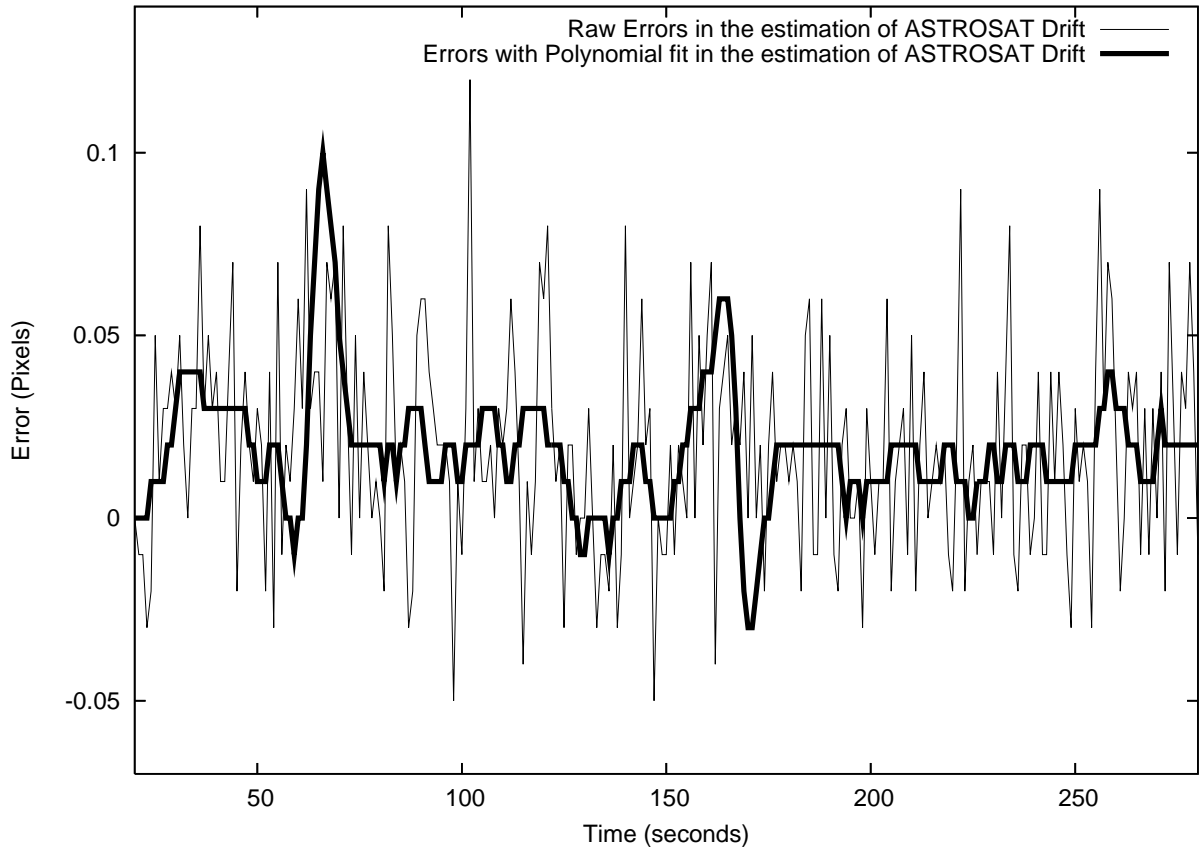}
\caption{Errors in the estimation of Satellite Pitch using visible channel of UVIT. Result is shown for the time interval 850 seconds to 1150 seconds as this interval provides the worst case conditions for the drift estimation. 
}
\label{fig-drift-error}
\end{figure}

From the simulated data on drift of the satellite(Figure~\ref{fig-drift}), a time interval (from 850 seconds to 1150 seconds) is chosen, corresponding to the duration when the X-ray Sky Monitor ( another payload on ASTROSAT ) is moved and the satellite is perturbed. This interval is assessed to be having the largest drift rates, and represent the worst case situation as far as recovery of the drift parameters is concerned. For a graphical illustration of the errors in the worst case, results for one of the fields from ESO HST Guide Star Cataloger {\it (with centre Galactic latitude = +90 degree and Galactic Longitude = 80 degree)} and for 10 photons for a magnitude 15 star, are shown in Figure~\ref{fig-drift-error}. It is evident from the plot shown in Figure~\ref{fig-drift-error} that even in the worst case the satellite drift can be determined using the observations from visible channel with accuracy $< 0.05$ CMOS pixels. With a pixel scale of $\sim$ 3.0 arc-sec per pixel, this corresponds to RMS error of $\sim 0.1$ arc-sec on the determination of satellite drift using visible channel time series data.

%
\section{Simulations to Generate Images from UVIT}
%

The aim of these numerical simulations was to generate the image frames as close as possible to the real observations including all the effects that could lead to deteriorate the final image quality and hence affect the photometric properties and angular resolution of the Telescopes. These include the effects of sky background, source intensity and saturation effects, Satellite Drift and Jitter, Optics performance, image frame acquisition rate, effects of background dark frames, and various detector parameters. While the final aim of the simulations was to study these effects on extended astronomical objects, images of artificial point sources were also simulated in order to understand these effects.

\subsection{Generating the UVIT Data Frames}

Data frames were simulated from input images of artificial points sources and extended sources (to be discussed in Section-5). The images for extended sources had input-pixel scale of 0.2 arc-sec and 0.5 arc-sec. Given total integration time and UVIT parameters (mirror diameter, CMOS pixel scale, frame acquisition rate etc.) each input-pixel would produce an average number of photons during the whole integration time. The actual photons in each input-pixel are then generated using Poisson Statistics. These photons are later randomly distributed in to total numbers of data frames. The blurring due to optics and detectors (i.e. due to spread of photoelectrons between photocathode and MCP) was approximated by a 2-dimensional Gaussian function with standard deviation ($\sigma$) of 0.7 arc-sec (on either axis) and spatial position of a photon in a data frame was determined by this function. The exact position of the photon in any data frame was recorded with accuracy of 1/8$^{th}$ of an input-pixel. Further the drifts of the satellite (Figure~\ref{fig-drift}) along Pitch, Roll and Yaw directions were incorporated so that each of the data frames were drifted with respect to each other. 
\par
To simulate UVIT photon counting detector, each of the photons were converted to a bunch of photoelectrons i.e a photon event. The number of photoelectrons in each bunch is obtained from a Gaussian distribution having 30000 photoelectrons as average with $\sigma$ of 6000 photoelectrons. These photoelectrons were then distributed over the face of the CMOS detector. The footprint of each of photon events (i.e the photoelectrons) were distributed over the area of 5 $\times$ 5 CMOS pixels following a 2-Dimension symmetric Gaussian probability distribution with $\sigma$ of 0.7 CMOS pixel. Figure~\ref{fig-events}A Shows footprint of a simulated photon event. Another Poisson distribution was invoked to obtain actual number of photo-electrons in a CMOS pixel. This number of photoelectrons were then divided by 20 to convert the photo-electrons in the units of counts. Hence a photon event  would correspond to a Gaussian distribution of average 1500 counts with sigma of 300 counts. Finally a randomly selected dark frame was added to each of the data frames. These dark frames were taken during the laboratory experiments of the UVIT detectors (Obtained from Mr. Joe Postma, University of Calgary, Canada, Private Communications). These dark frames were having random fluctuations in pixel values with RMS of $\sim$ 1 counts along with a gradient of $\sim$ 200 counts across the frame. Thus finally UVIT data frames of 512 $\times$ 512 CMOS pixels were generated containing the footprints of photon events against a laboratory recorded dark frame.

\subsection{Reconstruction of the Image: Centroid Finding Algorithms and Energy Thresholds}

The reconstruction of the image from UVIT data frames involves two critical steps: (i) Detection of the photon events within a data frame and (ii) Calculation of the event centroid for the detected events. While in actual practice these two tasks would be performed by the hardware arrangements on the payload itself, same strategy has been simulated here also. Photon events are detected by scanning the UVIT data frames and by applying one of the three event detection algorithms namely, 3-Cross, 3-Square and 5-Square. These algorithms scan the data frame and compare value of each pixel to its surrounding pixels. The definition of surrounding pixels varies for these algorithms. 3-Cross algorithm uses the adjacent single pixels along rows and columns from this central pixel; 3-Square algorithm uses the surrounding 8 pixels in the shape of 3 $\times$ 3 matrix, and 5-Square algorithm takes surrounding 24 pixels in the shape of 5 $\times$ 5 matrix as comparison. Each of the algorithms applies the following three criteria to detect a photon events in a data frame \citep{hutchings2007}:

\begin{itemize}
\item A pixel may be a candidate of an photon event centre if its value is larger than surrounding pixels contained in centroid window.
\item The value of this central pixel must exceed a minimum energy threshold to discard the fake events due to random variation in the background.
\item Another energy threshold is applied to the total energy of an photon event, which is the sum of all the pixel value in algorithm shape. The total energy of an photon event should also exceed this threshold. 
\end{itemize}

In case if two central pixels of a photon event candidate have equal values, it would not be considered as a photon event even if it is a genuine footprint. This would lead to a certain fraction of photons ($\sim0.6\%$) that would be missing in any case irrespective of the centroid algorithm used. Event centroids are later calculated using the centre of gravity method \citep{michel1997} following the various algorithm pixel configuration. The centroid coordinates are estimated up to much higher resolution than a CMOS pixel. The background level for a detected photon event is estimated from the minimum of 4 corner pixels of 5 $\times$ 5 pixels around the event centre.
\par
\begin{figure}
\centering
\plotone{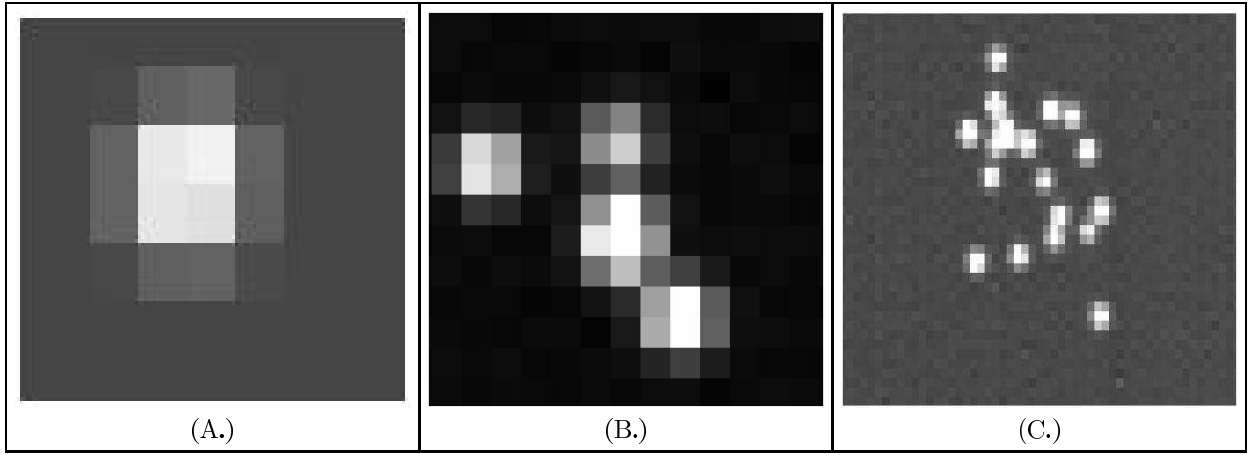}
\caption{Photon Event footprints on CMOS detector. (A) shows a Single Photon Event, (B) shows a configuration of few overlapping photon events and (C) depicts a section of UVIT data frame corresponding to a sky region of high count rate. The overlapping photon events would lead to wrong determination of the event centroids and may cause the incorrect number of detected events in a frame.
 }
\label{fig-events}
\end{figure}

Another important factor that would affect the imaging performance of the system is the presence of another photon event in the neighborhood of one event in a data frame. This situation is referred as {\it ``Double Photon Events''}. In situation of double photon events the footprints of two events would mix up and this would lead to incorrect estimate of the event centroid, thus affecting the angular resolution. Further if the two events are close enough one of them may remain undetected. Figure~\ref{fig-events}B shows couple of footprints of a simulated overlapping photon events. Figure~\ref{fig-events}C depicts the situation in a crowded field. These multiple overlapping photon events would be a major source of concern especially in case of extended sources and it is highly desirable to keep trace of such incidents. Double photon events can be identified by putting another threshold namely {\it Rejection Threshold}. The rejection threshold is the difference between the highest and lowest values of the 4 corner pixels of 5 $\times$ 5 pixels matrix around the event central pixel. In case of genuine single photon events, this difference would expected to be very low, however the presence of another photon event in the neighborhood would raise the counts in one of the corner pixels well above the background. Therefore a high value of the rejection threshold i.e. {\it ``Corner Maximum - Corner Minimum''} would be an indication of double photon event. 
\par
This is to mention that the statistics of detected events would be affected by the presence of double or multiple events.  While the percentage events detected by 3-Cross algorithm is highest, it is lowest for 5-Square algorithm. This is basically due to the fact that 3-Cross algorithm has the smallest footprint and thus least affected by overlapping events. It is also consistent with the laboratory investigation of the detectors \citep{hutchings2007}. However 3-Cross algorithm may also give rise to fake events detection, due to its shape. Given the shape of centroid window of 3-Cross algorithm one single isolated photon event could be detected as two photon events or two overlapping photon events could be detected as three photon events. In first case when a single isolated photon event falls near a corner of CMOS pixel, the neighboring 2 $\times$ 2 pixels would be getting nearly equal counts. If due to fluctuations the pixels that are diagonally opposite have counts greater than that are in rest of the two pixels, the 3-Cross algorithm would detect two events centred at the diagonally opposite pixels. In second case, two (or more) overlapping photon events could lead to fake event detection. Here two photon events (one of them falls near a corner of the pixel) overlaps along a diagonal in such a way so that they would appear as three photon events to 3-Cross algorithm. It is worth noting that neither 3-Square nor 5-Square algorithm would detect such fake events due to their extended footprints.

%
\subsection{Errors in Centroid Determination}
%

The issue of centroid errors in case of intensified CCD photon counting detectors has been discussed by several authors \citep{dick1989, michel1997, hutchings2007}. As discussed earlier the centroids of the photon events are calculated from centre of gravity method by three different algorithms using different configurations of pixels around the central pixel. In addition to random errors associated with the noise and background variation \citep{michel1997}, this method also put a systematic bias over the calculated values of the centroids in form of {\it ``Fixed Pattern Noise (FPN)''} \citep{dick1989, michel1997}. The FPN would be visible in the reconstructed images in a form of a regular grid structure, repeated at CMOS pixel scale, imposed on the images. There are various factors that can cause FPN \citep{dick1989}, however the main reason of FPN in reconstructed UVIT images is the exclusion of wings of footprint of a photon event by pixel configuration used by centroid algorithms. As the simulated photon event footprints have the FWHM of $\sim$1.65 CMOS pixel, the wings of the energy distribution would be left out by the 3-Square and 3-Cross algorithms while 5-Square shape would collect almost all the energy from a photon event. Therefore the FPN is clearly visible in the reconstructed images by 3-Square and 3-Cross centroid algorithms, whereas images generated by 5-square algorithm do not suffer from this effect.
\par
 A correction, to remove systematic bias (in form of FPN), has been done by comparing the theoretical and detected cumulative probability distribution of flat field centroids data. In absence of any systematic effect this distribution were expected to be a constant. This information was used to find the true centroid of the event by removing the systematic bias. For this purpose over 1 million photons events were simulated randomly on the face of the pixel and were detected by using all the three centroid algorithms. The decimal parts of the calculated centroid position were used to form the cumulative probability distribution. By comparing the probability density functions of both the theoretical and calculated curves, true centroids were determined for the calculated centroids data. This correction were applied to centroid data generated by 3-Cross and 3-Square algorithms only as systematic biases due to FPN were very prominent in these cases.
\par
\begin{figure}
\centering
\plotone{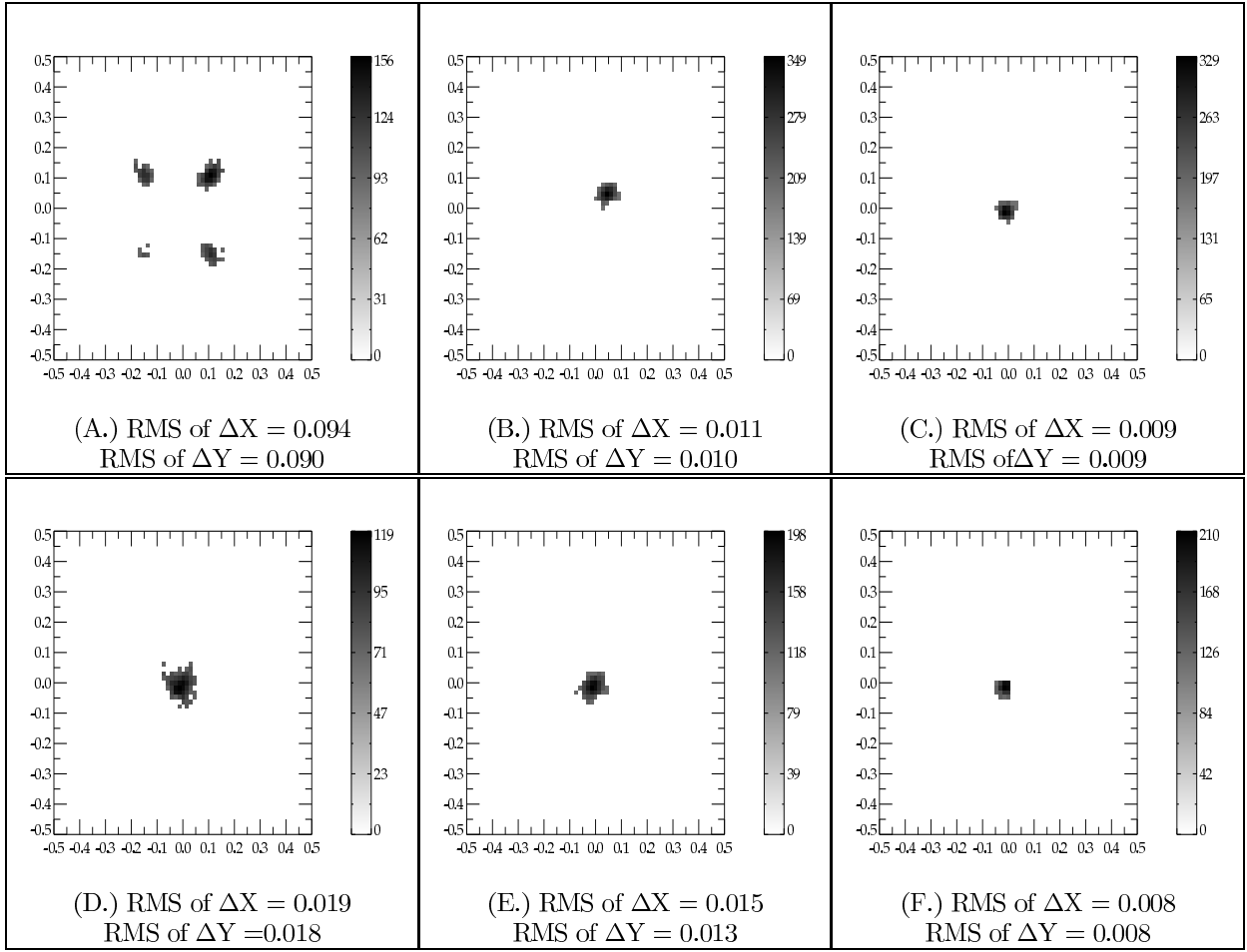}
\caption{Error maps for the 3-Square algorithm. These error maps represent the face of the CMOS pixel with centre as the original position of the incoming photon. A data point in these maps corresponds to detected position of the photon event with respect to original one. The upper panel (A, B and C) show the uncorrected data with systematic bias while the lower panel (D, E and F) corresponds to the data corrected to remove systematic bias. The left most plots (A and D) are for the case when photon is falling at a corner of the CMOS pixel. The central plots (B and E) corresponds to the case when photon is falling in between the centre and corner of the pixel along the diagonal line and the right most plots (C and F) are for the case when photon is falling at the centre of the pixel. The RMS value of the scatter are given along with the plots for each of the axes. (The threshold on central pixel energy used = 150 counts; The threshold on total event energy used = 450 counts)}
\label{fig-errormap-3sq}
\end{figure}

Along with this position dependent systematic bias, the random errors are also present in centroid estimation due to noise in the background dark frame used to record the photon event footprint. The dark frame used in the data simulation has a variance of $\sim$1 count along with a gradient across the frame of $\sim$ 200 counts. This leads to RMS error of $<0.01$ pixel in the calculated centroid position. To estimate the systematic bias and errors on the calculated centroid positions, photons events were simulated on the dark background frames. 1000 photon events were simulated corresponding to a location over the pixel face and their centroids were determined using each of the three centroid algorithms. In case of 3-Cross and 3-Square algorithms the necessary corrections were also applied to remove the systematic bias due to FPN. The actual position of the photon was known with the accuracy of 1/64$^{th}$ of a pixel, so by the comparing the input and output centroids the bias and errors were estimated. The bias and random errors in the determination of event centroids are shown in Figure~\ref{fig-errormap-3sq} in form of 2-D error maps for 3-Square algorithm. These error maps represent the face of a CMOS pixel with the centre being the actual position of the photon. A data point in the error maps represents the detected position of the photon event. Results for three locations on the CMOS pixel are presented; (i) Near a Corner (ii) at the midpoint of the centre and a corner along the diagonal and (iii) at the centre of the CMOS pixel. The effect of FPN is clearly visible in the case when the photon events were falling near the pixel corner (Figure~\ref{fig-errormap-3sq}A), putting a systematic bias of $\sim$0.15 CMOS pixel on the calculated position of the photon and causing the calculated centroids to fall in any one of the four corner pixels. It can also be seen in case when photon is falling between a corner and the centre of the pixel with bias of $\sim$0.05 pixel (Figure~\ref{fig-errormap-3sq}B). A shift of $\sim$ 0.01 pixels is noticed in all the cases which is due to precision limit of the simulation (1/64$^{th}$ of a pixel). The RMS of the scatter are between $\sim$0.01 to $\sim$0.02 pixels and is minimum for the case when the photon events were falling at the center of the pixel. The error maps in Figure~\ref{fig-errormap-3sq} suggest that the accuracy in the centroid determination would depend on the location of photon event over the pixel face and vary between  $\sim$0.01 to $\sim$0.02 pixels from centre of the pixel to corner of the pixel.

%
\section{Photometric Properties of the Reconstructed Images}
%

As explained in Section-3 the image reconstruction from UVIT data frames involves the use of various energy thresholds. It is anticipated that the photometric accuracy of the reconstructed images would depend on the choice of these thresholds. The values of energy thresholds for event detection could lead to systematic photometric bias over the pixel face for event detection. Also if the rejection threshold is not chosen properly the double photon events would be used for image reconstruction thus deteriorating the image resolution. Although the photometric non-linearity in reconstructed images were expected due to nature of photon counting detectors, the results from the simulations indicate that photometric non-linearity in reconstructed images follows a complex pattern which is closely tied up with values of thresholds used. All these effects are discussed in the following subsections.

%
\subsection{Photometric Variation over the Pixel due to Energy Thresholds}
%

 The distribution of energy of a photon event over the CMOS pixels would depend on the the location of the photon event over face of the pixel. The photon events falling in the centre of the pixel would contain a higher amount of their energy within the centroid algorithm shape as well as in the central pixel compare to the photon events falling near the edges/corners of the CMOS pixels. Therefore values of the energy thresholds for central pixel energy and total event energy would result in a selection bias over the face of a pixel and the probability of selection of photon events would not be uniform over the face of the pixel. This non-uniformity would also vary among all the three centroid algorithms given their different centroid windows shape.
\par
\begin{figure}
\centering
\plotone{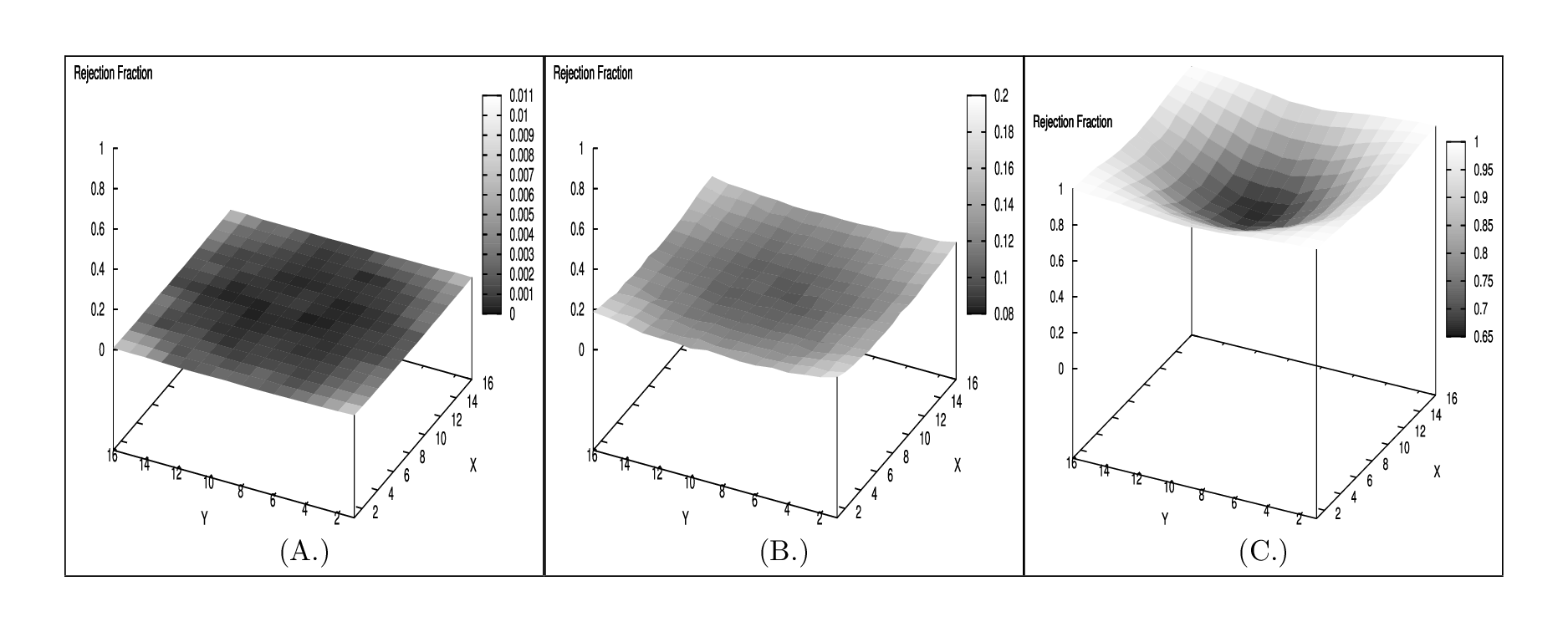}
\caption{Plots showing the non-uniformity over CMOS pixel face due to energy thresholds for 3-Square algorithm. These plots show the variation of the fraction of rejected photon events over the CMOS pixel, for different set of energy thresholds. Plot (A) represents the case when the central pixel energy threshold and the total energy threshold are kept at low values of 150 counts and 250 counts respectively. Plot (B) is for the case when total energy thresholds is given a high value of 1050 counts while keeping the central pixel energy at a low value of 150 counts. The plot (C) show the variation when central pixel energy is kept at high value of 450 counts while the total energy threshold is at low value of 650 counts.
}
\label{fig-pxlface}
\end{figure}

To estimate this non-uniformity, photon events were simulated on various locations over the pixel face. A pixel is divided in to 16 $\times$ 16 cells for this purpose. For the given the set of energy thresholds, the fraction of rejected events for each cell is determined. This rejection fraction over the face of the CMOS pixel is shown in Figure~\ref{fig-pxlface} for different set of energy thresholds and for the 3-Square algorithm. As shown in the plot (A), for low values of energy thresholds, the rejection fraction is negligible ($< 1\%$). Hence there would not be any selection bias or non-uniformity over the face of the CMOS pixel. Plot (B) represents the case when the threshold on central pixel energy is kept very low and threshold on total event energy is kept very high, in this case the selection of the event would be preliminary decided by the total energy threshold. Here the response for 3-Square algorithm is not uniform over the pixel face and $\sim 20\%$ of the events falling near the corner would remain undetected due to thresholds limit. This is consistent as for the events falling near a corner of CMOS pixel, a significant amount of the total event energy falls outside the 3-Square algorithm shape. On the other hand if the threshold on central pixel energy is given a very high value and threshold on total event energy is kept at low value, the non-uniformity is clearly visible and prominent over the pixel face. With a high threshold value for central pixel energy almost all the events falling on the corners of the pixel are rejected and a significant non-uniformity is observed varying between $65\%$ in the centre to $100\%$ on the corners.
\par
Thus the selection of energy thresholds decides the non-uniformity of event detection over the face of a pixel. It is interesting to point out that the other two algorithms would also be affected due to choice of energy thresholds. The 5-Square algorithm would collect most of the photon event energy contained within the footprint irrespective of its location over the pixel face, and thus would be less sensitive to the total energy thresholds. On the other hand the events falling near the corners of the pixel would sent a significant amount of their energy out of the 3-Cross algorithm centroid window, hence making it most sensitive to total energy threshold. The threshold on central pixel energy would affect all the centroid algorithms in same manner as the same central pixel energy is being used. Although low threshold values would result in to flat response over the pixel face but it would lead to detection of fake events too. On the other hand high thresholds would start rejecting the genuine events with an additional selection bias over the pixel. 

%
\subsection{Impact of Double Events over Photometric Linearity in Simulated UVIT images}
%
 
As discussed in Section-3.2, the double photon event is referred to a situation when two photon events fall very near to each other in a UVIT data frame, such that their footprints overlap. These double or multiple photon events in UVIT data frames are serious issue of concern. Although such events could be identified using the rejection threshold (discussed in Section-3.2), still they have significant effects over the photometry and angular resolution of the reconstructed images. Such events would affect the final image in various ways such as, deteriorating the angular resolution, inaccurate photometry, non-linearity in the reconstructed images etc. As will be discussed below, in case of sources having high count rate per frame, these double/multiple photon events would not only give rise to the photometric non-linearity but also lead to complicated patterns in reconstructed images. In order to understand the consequences of these effects in imaging of astronomical sources (such as Galaxies) that has complex structures, the nature of these effects has been studied for artificially simulated points sources.
\par
Point sources (having extent of 1/64 arc-sec $\times$ 1/64 arc-sec) with count rate 25 counts/sec, were simulated against the average sky background of 0.004 counts/sec/arc-sec$^2$. The data frames were generated for 3000 seconds with the frame acquisition rate of 30 frames per second. The effect of optics was not considered in these simulations. The modeled satellite drift had been added to each of the image frames. The same drift was later removed while reconstructing the final image but incorporating the errors in the drift corrections. Later the output images of these points sources were reconstructed using these data frames by all the three centroid algorithms, with scale of 1/32$^{th}$ arc-sec. The central pixel energy threshold of 150 counts and total energy threshold of 450 counts were used for event detection; with these values of thresholds the non-uniformity over the CMOS pixel face has been found to be $ < 1\%$ RMS. Also systematic bias due to Fixed Pattern Noise was corrected in case of 3-Square and 3-Cross algorithms, as discussed in Section-3.3. 
\par
The double/multiple photon events are responsible for non-linearity in photometry of reconstructed images. It turns out that the presence of a strong source would severely affect the photometry of neighboring regions, giving rise to photometric distortion in the reconstructed images. To quantify the consequences of such a case, the ratio of final reconstructed images to the corresponding input image (constructed with known positions of all the photons on the detectors, irrespective of single/double events) were taken. The ratio was taken after the smoothing of both the images through convolution from Gaussian functions. Both the images were convolved with two Gaussian functions of having standard deviations ($\sigma$) of 0.5 arc-sec and 0.25 arc-sec. The first Gaussian convolution was truncated up to $\pm 2\sigma$ on either side while next Gaussian was truncated after $\pm 3\sigma$. 
\par
\begin{figure}
\centering
\plotone{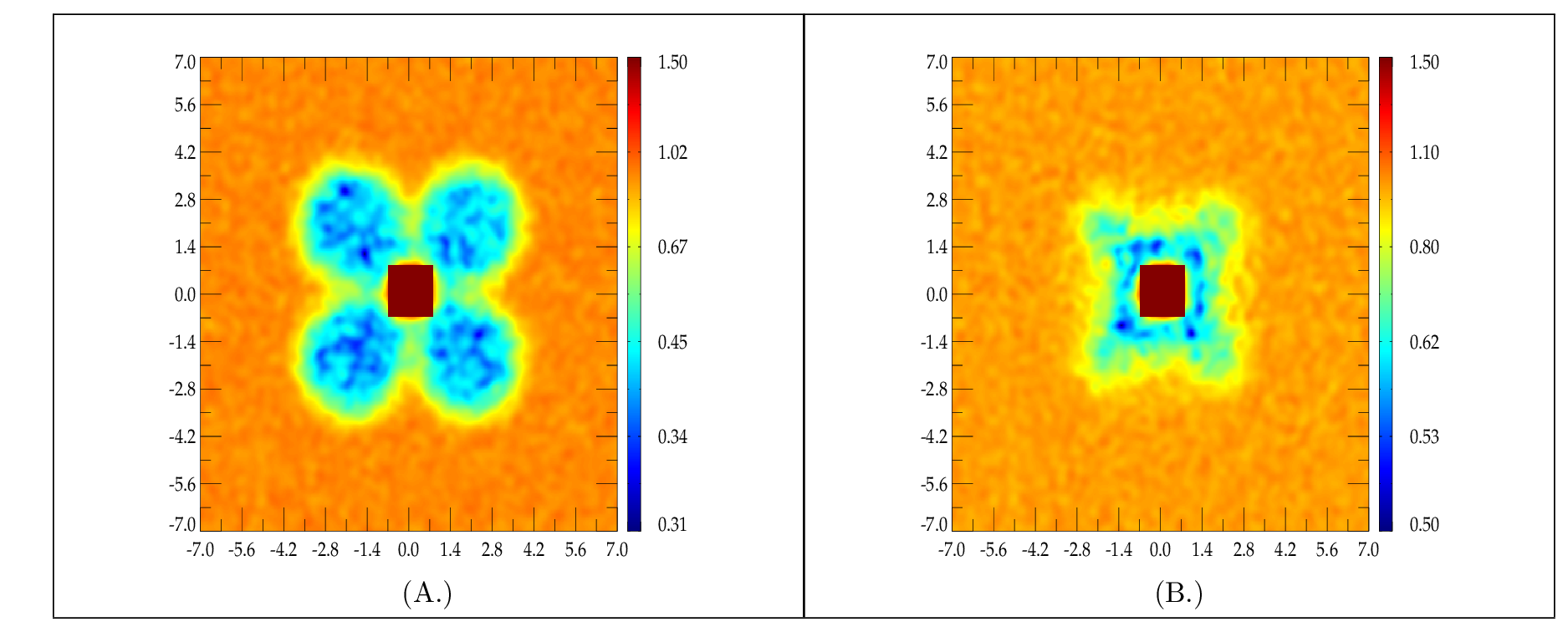}
\caption{Ratio maps showing the photometric distortion in the background due to presence of a strong point source for 3-Square algorithm. These maps are generated by taking the ratio of final reconstructed image to the image created with known positions of all the photons on the detector. The effects of optics has not been considered, thus these maps indicate the effects of detector parameters and satellite drift on image reconstruction. The errors in drift corrections were incorporated. Plot (A) is for rejection threshold of 40 counts and plot (B) corresponds to rejection threshold of 500 counts. The X and Y axes are in units of CMOS pixels. The central region of one CMOS pixel, containing the source has been masked. (The threshold on central pixel energy = 150 counts; The threshold on total event energy = 450 counts)
}
\label{fig-ratiomap-withdrift}
\end{figure}

Figure~\ref{fig-ratiomap-withdrift} show these ratio maps for 3-Square algorithm for the rejection threshold of 40 counts and 500 counts (see Section-3.2) respectively. In both the maps central one CMOS pixel area containing the source has been masked. These ratio images show very interesting and complex structure of the photometric properties of the reconstructed images. The reduction in background intensity has been observed to be as low as $\sim 30\%$. As the probability to loss a background photon due to another background photon is negligible; the presence of a strong source is main cause of such deficiency of the background photons. In fact it turns out that a strong source would affect the photometry of neighboring regions more severely compare to its own photometry. As an estimate, for the source count rate of 25 count per second, the probability that a background photon and source photon occur in same frame is $\sim 57\%$, whereas the probability of occuring two or more source photons in the same frame is $\sim 20\%$. Further it is clear from the ratio maps shown in Figure~\ref{fig-ratiomap-withdrift}, the rejection threshold used to indicate double photon events can also have a significant effect over such photometric structures, as this parameter is responsible for selecting the events that would be used for image reconstruction. In the first case when rejection threshold is kept at low value (40 counts) it would reject most of the possible double events. Since the constraint on the double photon events is strict it would also reject the barely overlapping events, thus extending the region of source effect. Whereas in the second case due to high value of rejection threshold (500 counts), it would select all the photon events, even the corrupted ones, thus reducing the extent of photometric structures.  
\par
It is found that the presence of double or multiple photon events in UVIT data frames could lead to photometric distortion in reconstructed images. Further it is shown that a point source with high count rate would give rise to complex structures in a uniform background in its neighborhood. Also it is anticipated that in case of extended sources these structures would be of even more complex nature, especially the neighborhood of hot spots in the source would be affected in very adverse way. This would be discussed in next section. 

%
\section{Photometric Effects on Simulated Extended Sources}
%

\begin{figure}
\centering
\epsscale{0.85}
\plotone{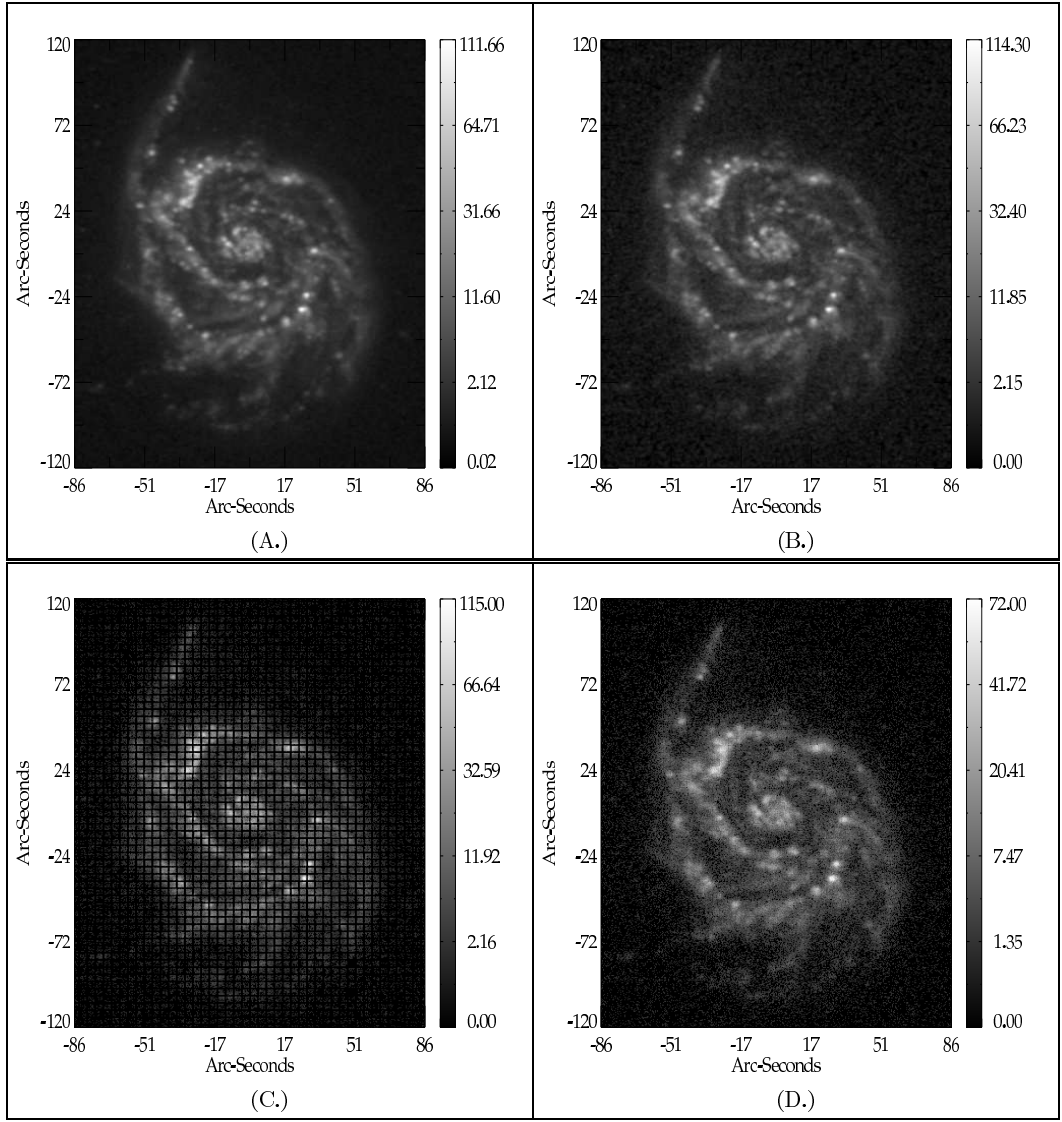}
\caption{The input and simulated images of a Galaxy. An archival Ultra-Violet image of M51 Galaxy from GALEX data archive is used for simulation. After applying necessary correction to the archival image an input image is obtained. This input image is shown in (A). (B) shows the simulated image of the input image, using Poisson statistics. This image is to be processed within UVIT subsystems. Both the images correspond to 3000 seconds integration time with the frame rate of 30 frame/second and are convolved with Gaussian function with $\sigma$ of 0.5 arc-sec. An uncorrected reconstructed image of the same galaxy using 3-Square algorithm is shown in (C). The superimposed grid pattern is due to Fixed Pattern Noise and has frequency of one CMOS pixel. The corresponding corrected image is shown in (D). The pixel scale in all the images is 0.5 arc-sec/pixel.
}
\label{fig-simulimages}
\end{figure}

All the effects discussed above give a detail picture of the characteristics of the reconstructed images, however the actual astronomical sources (e.g. galaxies etc.) are known to have more complex structures with various regions of varying intensities, background etc. Hence photometric accuracy and distortion present in these cases would also expected to be of complex nature. To explore the effects of above discussed factors on the reconstructed images of real astronomical sources, an Ultra-Violet image of M51 Galaxy from GALEX data archive\footnote{http://galex.stsci.edu/GR4} is used for the simulations. Necessary corrections are applied to pixel counts of this archival image such that the same galaxy is being observed by UVIT at a distance three times farther than the actual one. This is done to match the scale of this input image with UVIT scales. In this way one pixel of input image corresponds to 0.5 arc-sec of the sky. An simulated sky image is generated from this input image using Poisson statistics (as described in Section-3), to be further processed within the simulated UVIT subsystems. A comparison of input image and simulated image is shown in Figure~\ref{fig-simulimages}A and Figure~\ref{fig-simulimages}B. The integration time of 3000 seconds is used, with frame acquisition rate of 30 frames/sec. Both the images are convolved with Gaussian having $\sigma$ of 0.5 arc-sec for smoothing. The reconstructed images are shown in Figure~\ref{fig-simulimages}C and Figure~\ref{fig-simulimages}D (see Section-3.2). The impact of Fixed Pattern Noise is shown in Figure~\ref{fig-simulimages}C in the uncorrected reconstructed image using 3-Square algorithm. The grid structure with frequency of one CMOS pixel is visible here. The corresponding corrected image is shown in Figure~\ref{fig-simulimages}D.
\par
\begin{figure}
\centering
\plotone{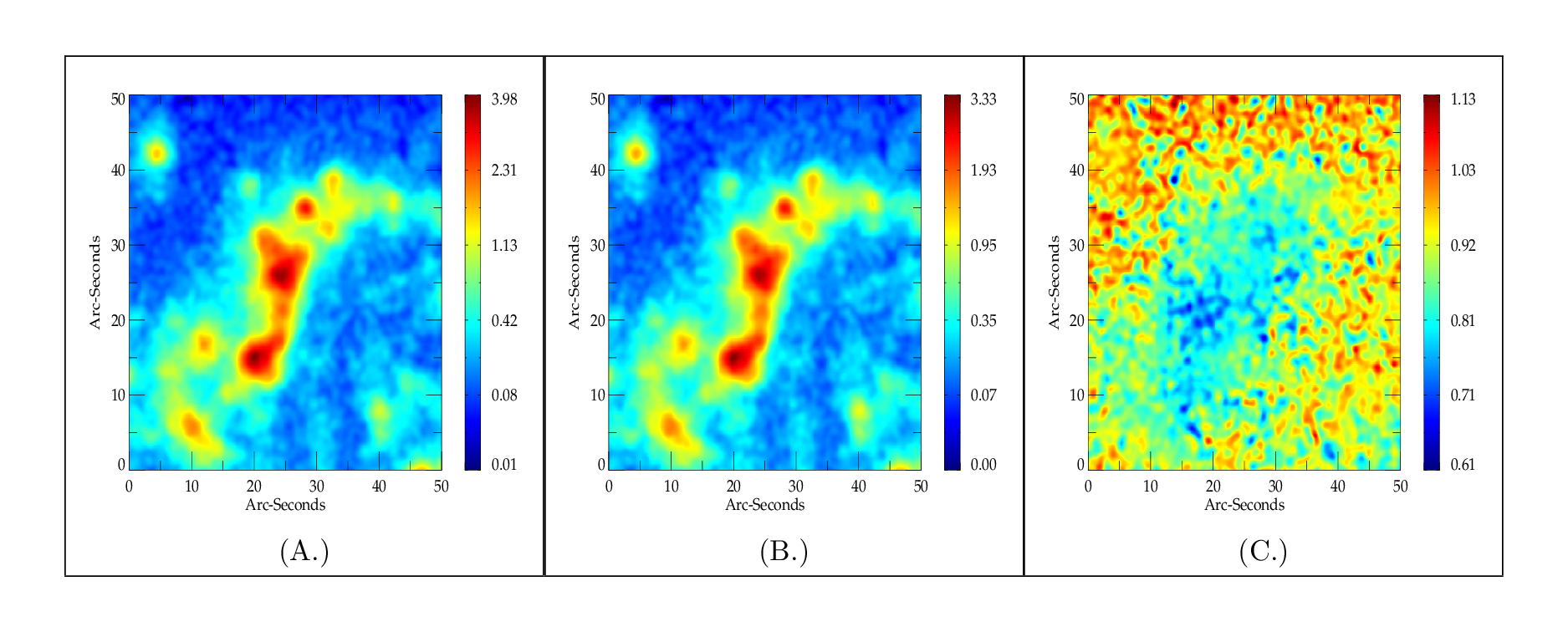}
\caption{Images showing the photometric distortion present in reconstructed image. A section of the image shown in Figure~\ref{fig-simulimages}. is used for this purpose. (A) is the image constructed using $''true''$ positions of the photon on the UVIT detectors, while (B) is the reconstructed image using 3-Square algorithm with rejection parameter of 40 counts. Both the images are convolved with Gaussian having  $\sigma$ of 0.5 arc-sec. (C) is ratio of images (B) and (A). The location of minima in the ratio image are away from the position of sources indicating a complex structure of photometric distortion in the reconstructed image. 
 }
\label{fig-ratio-gal}
\end{figure}

To estimate photometric accuracy and photometric distortion in the reconstructed image a section of galaxy's spiral arm (from the top left corner) is shown in Figure~\ref{fig-ratio-gal}. The image has been reconstructed with rejection parameter of 40 counts. Figure~\ref{fig-ratio-gal}A shows the reconstructed image with $''true''$ positions of the simulated photons, on the detectors. This represents the image if all the photons were detected irrespective of energy thresholds, double/multiple events etc. Figure~\ref{fig-ratio-gal}B is the reconstructed image with detected photon events using 3-Square algorithm. The pixel scale is $1/8^{th}$ arc-sec in both the images. For smoothing purpose both the images are convolved with Gaussian having $\sigma$ of 0.5 arc-sec. The effect of strong sources on the neighboring background is estimated by taking the ratio of images in Figure~\ref{fig-ratio-gal}B and Figure~\ref{fig-ratio-gal}A and is shown in Figure~\ref{fig-ratio-gal}C. The effect of a strong source on the surrounding background (as discussed in Section-4) is again visible here. The ratio image in Figure~\ref{fig-ratio-gal}C indicates a complex photometric structure in the reconstructed image. The regions away from the source show good recovery of photon events (as the ratio is $\sim1$) leading to accurate photometry. However, the nereby regions around the source suffer from loss of photon events and these are the regions where the photometry is most corrupted. For the section of galaxy image shown in  Figure~\ref{fig-ratio-gal}, this loss is as high as $\sim 40\%$. It is to be noted that the regions of inaccurate photometry are not symmetric around the source, indicating that intensity profile of the source could generate photometric distortion around its surrounding in a complicated shape. Also the source itself suffers from the photometric saturation effects and only $\sim 80\%-90\%$ of the source photons are recovered. The large fluctuations in the ratio image away from the sources is due to small number statistics as the sky background is extremely low ($\sim 10^{-5} photons/sec/arc-sec^2$).  Though both the images (Figure~\ref{fig-ratio-gal}A and Figure~\ref{fig-ratio-gal}B) contain roughly same number of photons in the background away from the source; a small error in photon's actual position would cause large fluctuations in the ratio image (Figure~\ref{fig-ratio-gal}C).

%
\section{Angular Resolution of the Simulated Images}
%

The angular resolution of the UVIT images is estimated by simulating the point sources as described in Section-4.2 but with incorporating all the known effects such as optics, satellite drift etc. that could deteriorate the image quality. It is found that the angular resolution of the reconstructed images is heavily dominated by the Point Spread Function (PSF) of the optics and spread in the trajectories of photoelectrons between photocathode and MCP. The other factors like, error in the satellite drift correction, errors in centroid estimation centroid estimation etc. only have a small effect. A 2-D Gaussian fit to the PSF profile yields $\sigma$ to be $\sim$0.7 arc-sec on either axis and the PSF appears to be independent of the choice of centroid algorithm and rejection threshold. $\sigma$ of the output images does not differ much from the input value of 0.7 arc-sec. This is consistent with the errors of centroiding (as discussed in Section-3.3) and errors in the satellite drift correction (see Section-2). However double/multiple photon events could change the profile of simulted PSF if the source is of high count rate per frame. It is found that an average count rate of 2 counts per frame could reduced the $\sigma$ of PSF up to $< 0.5 arc-sec$.
\par
\begin{figure}
\centering
\plotone{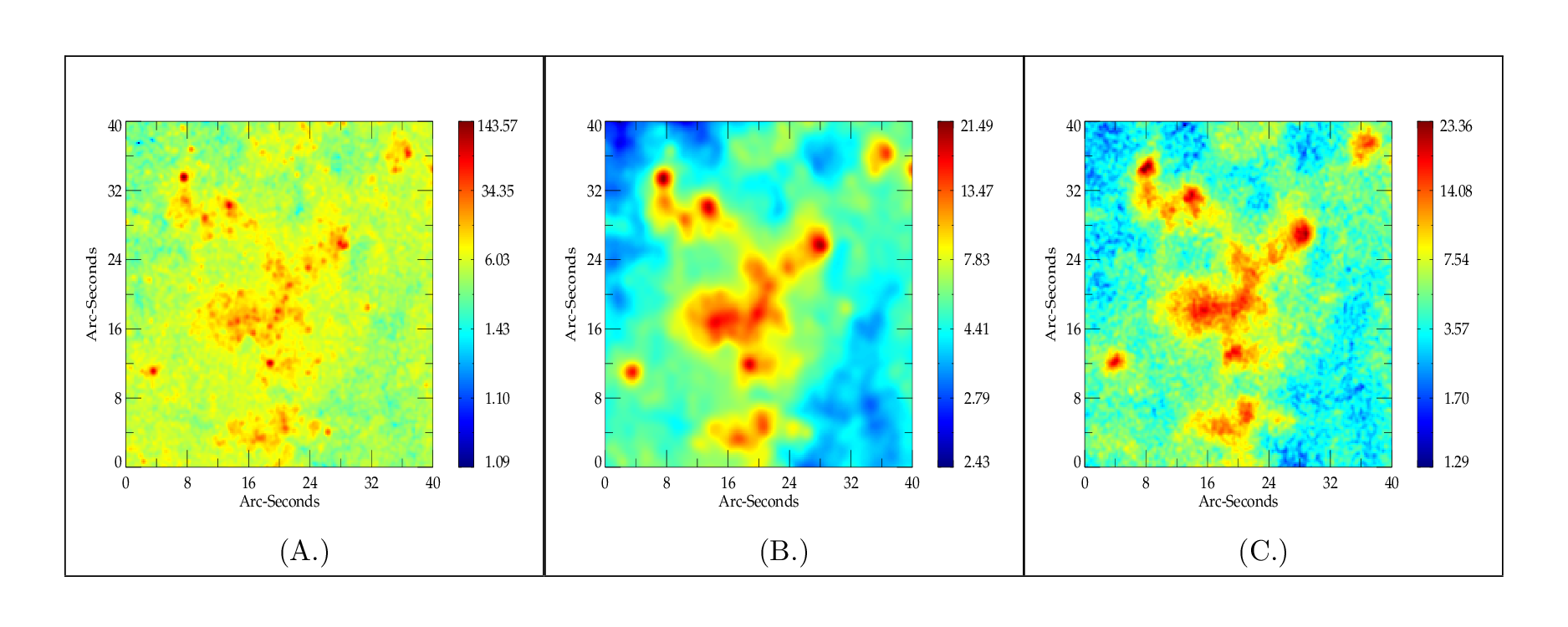}
\caption{Input and Simulated images of a portion of the Galaxy's spiral arm, showing the effect of UVIT Point Spread Function on the image. (A) is the input Hubble ACS image in B band. This image is convolved with PSF of the UVIT. The resultant image is shown in (B). The reconstructed image using 3-Square algorithm with Rejection Threshold of 500 is shown in (C). The pixel scale is 0.05 arc-sec/pixel and the integration time is 6000 seconds. All the images are convolved with Gaussian function having $\sigma$ of 0.2 arc-sec for smoothing purpose.}
\label{fig-angres-gal}
\end{figure}

To see the effect on extended sources an Hubble ACS B band image\footnote{Taken from HST data archive: http://archive.stsci.edu/hst} of the same galaxy (as used in Section-5) is simulated with pixel scale of 0.05 arc-sec per pixel. The input and output reconstructed images of a portion of the galaxy's spiral arm are shown in Figure~\ref{fig-angres-gal} for integration time of 6000 seconds. It is evident that the sources separated by $\sim 3.0$ arc-sec can easily be resolved in the reconstructed image. The convolution of input image with the simulated PSF of the UVIT is also shown for the comparison. The reconstructed and convolved images looks very similar to each other conforming the effect of PSF on the reconstructed images.

%
\section{Summary}
%

The results presented in this work are based on the numerical simulations performed to evaluate the expected performance of the Ultra-Violet imaging Telescope (UVIT). It is shown that the imaging characteristics of UVIT would be affected by a number of factors. Apart from performance of the optics, the other important factors are the drift of the satellite and functioning of Photon counting detectors. The final image is to be reconstructed with the UVIT data frames using some centroid algorithms. Also the satellite drift has to be removed from the UVIT data frames before using them for image reconstruction. To estimate the satellite drift, simulations have been done depicting the observations of some sky fields containing point sources through the visible channel of the UVIT. The centroids of the sources in the field are used to track the satellite drift. The results from these simulations indicate that drift of the satellite could be recovered with an accuracy of $< 0.1 arc-sec$.
\par
The simulations of UVIT data frames have been done including all the known effects of satellite drift, optics and photon counting detectors to study the photometric properties of the system. The centroids of photon events can be determined by three different algorithms (namely 5-Square, 3-Square and 3-Cross) with different pixels shape using centre of gravity method. The fluctuations in the background dark frames and in photon event footprint could result error in centroid estimations. These errors are estimated to be $\sim 0.01$ CMOS pixel or $\sim 0.03 arc-sec$. Also the effects of well known {\it Fixed Pattern Noise} has also been seen in the centroid determination process and in the reconstructed images in case of 3-Square and 3-Cross algorithms. It has been corrected before reconstructing the image.
\par
The characterictics of reconstructed images from UVIT observations and effects of various thresholds on image reconstruction are studied by simulating the archival images of a galaxy from GALEX and Hubble ACS data archives. As a galaxy (or other extended astronomical sources in general) is of much complex nature, several artificial point sources are also simulated to explore these effects. Simulations with artificial point sources show that the photometry of the reconstructed images would depend on the choice of two energy thresholds used by centroid algorithms to detect photon events in the UVIT data frames. For given energy thresholds events falling in the centre of CMOS pixel would be more probable to detect than the events falling near the corners of the pixel. By choosing suitable energy thresholds one could reduce such variations to $< 1\%$ over the pixel face. Further the photometric accuracy of the reconstructed images would be affected due to appearance of double or multiple photon events, which occur with a significant probability in and around bright sources. These photometric effects could be serious issue of concern for bright sources. It turns out that presence of a strong source could also produce complicated photometric patterns in the surrounding. These patterns depend on the choice of the rejection threshold. The photometric distortion in the reconstructed images of galaxy is of complex nature and depends on the intensity profile of bright sources present in the galaxy. The regions away from such bright sources offer nearly accurate photometry while the nereby regions of such sources suffer from higher photometric inaccuracies. The results of simulations shows that the photometric accuracy in these regions can be as low as $\sim 60\%$. Also the strong sources in the galaxy themselves suffer from photometric saturation effects, causing the recovery of $\sim 80\%-90\%$ of total photon events.
\par
 The PSF of the reconstructed images follows a 2-D Gaussian profile with $\sigma$ of 0.7 arc-sec on either axis and is dominated by the performance of the optics and drift of photoelectrons between the photocathode and MCP.  Further the PSF is found to be same in case of different centroid algorithms and different rejection thresholds. However the presence of double/multiple photon events could also cause PSF profile to vary in case of bright sources. The simulations of a Hubble ACS image shows that sources separated by 3.0 arc-sec are clearly resolved in the reconstructed images.
\par 
It is of interest to point out that the results of this study are not limited to performance of UVIT but are also applicable to any CCD or CMOS readout based photon counting detectors in general, which works on a fixed read out rate of frames.


\acknowledgements
Based on observations made with the NASA Galaxy Evolution Explorer. GALEX is operated for NASA by the California Institute of Technology under NASA contract NAS5-98034. Some of the data presented in this paper were obtained from the Multimission Archive at the Space Telescope Science Institute (MAST). STScI is operated by the Association of Universities for Research in Astronomy, Inc., under NASA contract NAS5-26555. Support for MAST for non-HST data is provided by the NASA Office of Space Science via grant NAG5-7584 and by other grants and contracts. We are thankful to Joe Postma (University of Calgary, Canada) for providing us necessary software tools for centroid determination and experimental data of UVIT detectors. We also thanks Shri Harish of ISRO Satellite Centre (ISAC), Bangalore for the simulation data of satellite drift. MKS thanks Council for Scientific and Industrial Research (CSIR), India, for the research grant award F.NO.9/545(25)/2005-EMR-I.

{}


\clearpage

\clearpage

\clearpage

\clearpage

\clearpage

\clearpage

\clearpage

\clearpage

\clearpage

\begin{thebibliography}{}

\bibitem[Bellis et al.(1991)]{bellis1991} Bellis, J.~G., Bone, 
D.~A., \& Fordham, J.~L.~A.\ 1991, \pasp, 103, 253 

\bibitem[Dick et al.(1989)]{dick1989} Dick, J., Jenkins, C., 
\& Ziabicki, J.\ 1989, \pasp, 101, 684

\bibitem[Hutchings et al.(2007)]{hutchings2007} Hutchings, J.~B., 
Postma, J., Asquin, D., \& Leahy, D.\ 2007, \pasp, 119, 1152 

\bibitem[Jelinsky et al.(2003)]{jelinsky2003} Jelinsky, P.~N., et 
al.\ 2003, \procspie, 4854, 233 

\bibitem[Michel et al.(1997)]{michel1997} Michel, R., Fordham, J., 
\& Kawakami, H.\ 1997, \mnras, 292, 611 

\bibitem[Sahnow(2003)]{sahnow2003} Sahnow, D.~J.\ 2003, \procspie, 
4854, 610 

\bibitem[Siegmund(1999)]{siegmund1999} Siegmund, O.~H.~W.\ 1999, 
Ultraviolet-Optical Space Astronomy Beyond HST, 164, 374 


\end{thebibliography}
\end{document}